\documentclass{article}

\usepackage{amssymb,amsmath}
\usepackage{times}

\usepackage{graphicx}

\usepackage{amssymb}

\begin{document}

\centerline{\Large{On correlation approach to scattering in the
decoherence timescale}}

\vspace{0.3cm} \centerline{\large{Towards the theoretical
interpretation of NCS and ECS experimental findings}}

\vspace{1cm} \centerline{C. Aris Chatzidimitriou-Dreismann$^1$ and
Stig Stenholm$^2$ }

\vspace{1cm} \noindent $^1$ Institute of Chemistry, Sekr C2,
Technical University Berlin, D-10623 Berlin, \\Germany.
\textit{Email:} dreismann@chem.tu-berlin.de

\vspace{0.3cm} \noindent $^2$Department of Physics, Royal
Institute of Technology, SE-10691 Stockholm, Sweden.
\textit{Email:} stenholm@atom.kth.se

\begin{abstract}
We provide a "first principles" description of scattering from
open quantum systems subject to a Lindblad-type dynamics. In
particular we consider the case that the duration of the
scattering process is of similar order as the decoherence time of
the scatterer. Under rather general conditions, the derivations
lead to the the following new result: The irreversible
time-evolution may cause a reduction of the system's transition
rate being effectuated by scattering. This is tantamount to a
shortfall of scattering intensity. The possible connection with
striking experimental results of neutron and electron Compton
scattering from protons in condensed matter is mentioned.
\end{abstract}


\section{Introduction}
The counter-intuitive phenomenon of entanglement \cite{QE}
 between two
or more quantum systems has emerged as the most emblematic feature
of quantum mechanics.  Experiments investigating entnaglement,
however, are mainly focused on collections of few simple (two- or
three-level) quantum systems thoroughly isolated from their
environment (e.g., atoms in high-$Q$ cavities and optical
lattices). These experimental conditions are necessary due to the
decoherence of entangled states. In short, decoherence refers to
the suppression of quantum superpositions caused by the
environment.
 By contrast, entanglement in condensed and/or molecular
matter at ambient conditions is usually assumed to be
experimentally inaccessible. However, two new scattering
techniques operating in the sub-femtosecond time scale provided
results indicating that short-lived entangled states may be
measurable in condensed matter even at room temperature
\cite{PRL97,highlights}.

In this paper we provide a first-principles treatment of
scattering from "small" open quantum systems in condense-matter
environments, in the "time window" of decoherence  of the
scattering system. That is, the focus is in "fast" scattering
processes with a duration (usually denoted scattering time,
$\tau_{sc}$) of the order to the scatterer's decoherence time,
$\tau_{dec}$. This may be considered to represent an "extension"
of standard scattering theory --- as applied e.g. to neutron
physics \cite{vHove,Squires} or electron scattering \cite{Weigold}
--- in which the concepts of entanglement and decoherence play
essentially no role. The first part of the derivations are
analogous to the standard (often denoted) "van Hove formalism"
\cite{vHove}; see also the textbook \cite{Squires}. Then a reduced
open quantum system, i.e. a micro- or mesoscopic system
characterized by a set of preferred coordinates, is introduced.
This corresponds to the "small" physical system that scatters a
neutron (electron, etc.) with a sufficiently large momentum
transfer. Its dynamics is described by a simple Lindblad-type
master equation \cite{Lindblad,Stig1} (which, for the sake of
simplicity, contains only one Lindblad operator, $X$), thus
including explicitly the effect of decoherence into the formalism.

The striking result of the derivation may be summarized as
follows: The irreversible time-evolution (owing to the Lindblad
operator $X$) may cause a reduction of the transition rate of the
system (from its initial to its final state). In "experimental"
terms, this is tantamount to an effective reduction of the
system's cross-section density and thus a shortfall of scattering
intensity.

\section{Scattering in brief}

We assume an N-body Hamiltonian $H_{total}=H_0+V$ with an
interaction of the form
\begin{equation}
V(\mathbf{r})=\lambda \,n(\mathbf{r})  \label{a1}
\end{equation}
where $\,n(\mathbf{r})$ is the particle density operator and
$\lambda $ is the rest of the interaction (contact potential). For
example, in the case of neutron scattering from a system
consisting of $N$ particles with the same scattering length $b$
one may put
\begin{equation}
n(\mathbf{r}) = \frac{1}{V}\sum_j \delta(\mathbf{r}-\mathbf{R}_j)
\label{a1-1}
\end{equation}
where $V$ is the volume, $\mathbf{R}_j$ is the spatial position of
the $j$-th particle, and
\begin{equation}
\lambda = \frac{2 \pi \hbar^2}{m}\, b \  ,
\end{equation}
$m$ being the neutron mass. (For further details about scattering
from "bound" and "free" particles, see the textbook
\cite{Squires}.)

In the \textit{interaction} picture, the Schr\"odinger equation is
now (putting for simplicity $\hbar =1$)
\begin{equation}
i\partial _{t}\Psi =\lambda\, n(\mathbf{r,}t)\Psi
\end{equation}
with the perturbative solution
\begin{equation}
\Psi (t)=\Psi (0)-i\lambda \int_{0}^{t}n(\mathbf{r},t^{\prime
})dt^{\prime }\Psi (0).  \label{a2}
\end{equation}

We write the transition probability $W(t)$ between initial states
$\psi_{i}$ (with probability $P_{i}$ that the scattering system is
in the state $\psi_{i}$) and final states $\psi _{f}$  of the
scattering system to be given by
\begin{equation}
W(t) = \sum_{i,f}\mid \langle \psi _{f}\mid \lambda
\int_{0}^{t}n(\mathbf{r}, t^{\prime })dt^{\prime }\mid \psi
_{i}\rangle \mid ^{2}P_{i}.
  \label{a3}
\end{equation}
It should be noted that $\psi_{i}$ and  $\psi _{f}$ are
eigenstates of the unperturbed N-body Hamiltonian $H_0$
\cite{Squires,vHove}. This allows us to write the transition
probability in the form
\begin{equation}
W(t) = \lambda ^2\int_0^tdt^{\prime }\int_0^tdt^{\prime \prime}
 \sum_f \langle \psi _f\mid n(\mathbf{r,}t^{\prime })\,
 \rho\, n(\mathbf{r,}
t^{\prime \prime })\mid \psi _f\rangle ,  \label{a3+1}
\end{equation}
where
\begin{equation}
\rho =\sum_i\mid \psi _i\rangle P_i\langle \psi _i\mid ,
 \label{rho}
\end{equation}
by noting that $n^\dag(\mathbf{r},t) = n(\mathbf{r},t)$, since
$\mathbf{R}_j$ and $\mathbf{r}$ are Hermitian operators.

In an actual scattering experiment from condensed matter, we do
not measure the cross-section for a process in which the
scattering system goes from a specific initial state $\psi _{i}$
to another state $\psi _{f}$, both being  unobserved states of the
many-body system. Therefore, one takes an appropriate average over
all these states \cite{Squires,vHove}, as done in Eq.~(\ref{a3}).

Furthermore, the initial ($\mathbf{k}_0$) and final
($\mathbf{k}_1$) momenta of an impinging probe particle (neutron)
may be assumed to be well defined \cite{Squires,vHove}.
  Introducing the momentum transfer $\mathbf{q}= \mathbf{k}_0 -
\mathbf{k}_1$ from the probe particle  to the scattering system,
 the Fourier transform  of the particle density
reads
\begin{equation}
 n(\mathbf{r},t) = \frac{1}{(2\pi)^3}
 \int\!d\mathbf{q}\, n(\mathbf{q},t)
 \exp\{i\,\mathbf{q}\cdot\mathbf{r}\}
\end{equation}
where, in the case of neutron scattering, cf.~Eq.(\ref{a1-1}),
\begin{equation}
n(\mathbf{q},t)= \sum_j \exp\{-i \mathbf{q}\cdot \mathbf{R}_j(t)
\} \ .
\end{equation}
Since $n(\mathbf{r},t) $ is Hermitian it holds
 $
n^\dag(\mathbf{q},t) = n(-\mathbf{q},t)
 $
and one obtains from Eq.~(\ref{a3})
\begin{equation}
W(t) = \lambda ^2\int_0^tdt^{\prime }\int_0^tdt^{\prime \prime}
 \sum_f \langle \psi _f|
n(\mathbf{q},t^{\prime })\,\rho \,n(-\mathbf{q},t^{\prime \prime
})
 |\psi _f\rangle ,  \label{a4}
\end{equation}

At this stage one traditionally assumes  that the sum over $\psi
_f$ runs over all possible eigenstates of $H_0$ which constitute a
complete set, i.e. $\Sigma_f |\psi _f\rangle\langle \psi _f|
=\mathbf{1}$; see \cite{Squires,vHove}. Hence
\begin{equation}
  \sum_f \langle \psi _f|
n(\mathbf{q},t^{\prime })\,\rho \,n(-\mathbf{q},t^{\prime  \prime
})
 |\psi _f\rangle =
 Tr\left[ n( \mathbf{q},t^{\prime })\,\rho
\,n(-\mathbf{q},t^{\prime \prime })\right]
  \label{a-ignore}
\end{equation}
 where $Tr[...]$ denotes the trace operation.
 As done in standard theory \cite{Squires,vHove},  in
Eq.~(\ref{a-ignore}) one
  first sums over all final states, keeping the initial state
$\psi _{i}$ fixed, and then  averages over all $\psi_{i}$ (see
e.g. \cite{Squires}, p.~19). The right-hand-side of
Eq.~(\ref{a-ignore}) contains the density operator $\rho$ of the
system before collision, Eq.~(\ref{rho}), which is a well known
result.


   By introducing a measurement time (the so-called
scattering time)  $\tau_{sc}$, that is the duration of the
scattering process, we find
\begin{eqnarray}
W(\tau_{sc}) &=&\lambda ^2\int_0^{\tau_{sc}}dt^{\prime
}\int_0^{\tau_{sc}} dt^{\prime \prime }Tr\left[ n(
\mathbf{q},t^{\prime })\,\rho \,n(-\mathbf{q},t^{\prime \prime
})\right]
 \nonumber\\
&=&\lambda ^2 \tau_{sc} \int_0^{\tau_{sc}}d\tau \,Tr\left[
n(\mathbf{q},t^{\prime })\,\rho \,n(-\mathbf{q},t^{\prime } + \tau
)\right] ,
 \label{a5}
\end{eqnarray}
where the stationary property of the correlation function has been
used \cite{Squires}. Now one can introduce the transition rate,
say $\dot{W}$, which is defined as
\begin{eqnarray}
\dot{W} \equiv \frac{W(\tau_{sc})}{\tau_{sc}}
 &=&\lambda
^2\int_0^{\tau_{sc}}d\tau \,Tr\left( n(\mathbf{k,} t^{\prime
})\,\rho \,n(-\mathbf{k,}t^{\prime }+\tau )\right)
   \nonumber\\
 &\equiv& \lambda
^2\int_0^{\tau_{sc}}d\tau \,C(\mathbf{q},\tau ).
 \label{a6}
\end{eqnarray}
Here the correlation function
\begin{equation}
C(\mathbf{q},t)= Tr[n(\mathbf{q}, 0)\,\rho \,n(-\mathbf{q},t)]
 \label{a7}
\end{equation}
is introduced, which is analogous to the so-called intermediate
function of neutron scattering theory \cite{Squires}.

\section{Irreversible dynamics}

We now introduce a set of preferred coordinates $\{\, | \xi
\rangle\}$, cf.~\cite{Zeh,Zurek}. These are the relevant degrees
of freedom coupled to the neutron probe. The density matrix needed
in (\ref{a5}) is then the \textit{reduced} one in the space
spanned by these states, and it is obtained by tracing out the
(huge number of the) remaining degrees of freedom belonging to the
"environment" of the microscopic scattering system (e.g. a proton
and its adjacent particles). To simplify notations, we denote this
reduced density matrix by $\rho$ too.

 In the
\textit{subspace} spanned by the preferred coordinates (also
denoted 'pointer basis'), we assume the relevant density matrix to
obey a Lindblad-type equation of the form \cite{Lindblad,Stig1}
\begin{equation}
\partial _t\rho =-i\left[ H,\rho \right] +\mathcal{R}\rho \equiv \mathcal{L}%
\rho  \label{b1}
\end{equation}
with the formal solution
\begin{equation}
\rho (t)=e^{\mathcal{L}t}\rho (0).
\end{equation}

\ Let us look at a time-dependent expectation value
\begin{equation}
\langle A(t)\rangle \equiv Tr\left( \rho (t)A\right) =Tr\left( e^{\mathcal{L}%
t}\rho (0)A\right) =Tr\left( \rho (0)e^{\mathcal{L}^{\dagger
}t}A\right) \  ,
\end{equation}
 where we define $\mathcal{L}^{\dagger }$ by setting
\begin{equation}
Tr\left( \left( \mathcal{L}X\right) Y\right) =Tr\left( X\left( \mathcal{L}%
^{\dagger }Y\right) \right) \  .
\end{equation}
 Thus we obtain a Lindblad time
evolution for the operators too by writing
\begin{equation}
\partial _tA(t)=\mathcal{L}^{\dagger }A(t).
\end{equation}
This form was actually the original Lindblad result. Note that
this works as long as $\mathcal{L}$ does not depend on time. For
time-dependent generators of the evolution, a somewhat more
elaborate scheme is needed.

Now we find that we may use this formalism to calculate
correlation functions like the one in (\ref{a6}). We write
\begin{equation}
\langle A(t)B\rangle =Tr\left[ \rho (0)\left(
e^{\mathcal{L}^{\dagger
}t}A\right) B\right] =Tr\left[ Ae^{\mathcal{L}t}\left( B\rho (0)\right) %
\right] \equiv Tr\left( A\rho _B(t)\right) ,
  \label{b2}
\end{equation}
where $\rho _B(t)$, as defined in Eqs.~(\ref{b2}), obeys the
equation
\begin{equation}
\partial _t\rho _B(t)=\mathcal{L}\rho _B(t)
\end{equation}
 and the initial condition
\begin{equation}
\rho _B(0)=B\rho (0).
\end{equation}
 Thus, except for the initial condition, we
have to solve the same equation of motion as for the density
matrix (\ref{b1}).

\section{Application to scattering}

We here assume a simple Lindblad-type ansatz for the master
equation in the relevant subspace. We set
\begin{equation}
\partial _t\rho =-i\left[ H,\rho \right] -K\left[ X,\left[ X,\rho \right] %
\right] =\mathcal{L\rho },
 \label{c1}
\end{equation}
where $K>0$ and $H$ is the reduced (or relevant Hamiltonian) of a
microscopic or mesoscopic scattering system and the double
commutator term describes decoherence. For simplicity of the
further calculations, we here assume that
\begin{equation}
\begin{array}{lll}
H\mid \xi \rangle & = & \mathcal{E}_\xi \mid \xi \rangle \\
&  &  \\
X\mid \xi \rangle & = & \xi \mid \xi \rangle .%
\end{array}
\label{b3}
\end{equation}
With Eq.~(\ref{a7}) we have
\begin{eqnarray}
C(\mathbf{q},\tau )
 &= &Tr [ n(\mathbf{q},0)\,\rho
\,n(-\mathbf{q},\tau )]
 = Tr\left( n(\mathbf{q},0)\,\rho
e^{\mathcal{L}^{\dagger }t}n(-\mathbf{q}, 0)\,\right) \nonumber \\
 & =& Tr\left( n(-\mathbf{q},0)\,e^{\mathcal{L}t}\left( n(\mathbf{q},
0)\,\rho \right) \right) \ .
\end{eqnarray}
This is equivalent with the expression
\begin{equation}
 C(\mathbf{q},\tau
)=\sum_{\xi ,\xi ^{\prime }}\langle \xi \mid n(-\mathbf{q},0)\mid
\xi ^{\prime }\rangle \langle \xi ^{\prime }\mid \rho _n(t)\mid
\xi \rangle .
\end{equation}

With the equation (\ref{c1}), one easily finds the well known
solution
\begin{eqnarray}
\langle \xi ^{\prime }\mid \rho _n(t)\mid \xi \rangle
 &=&
\exp\left[ -i\left( \mathcal{E}_{\xi ^{\prime }}-\mathcal{E}_\xi
\right)
 t  \right]
 \exp \left[ -K\left( \xi ^{\prime }-\xi \right) ^2t\right]
 \nonumber\\
&  &\times \ \langle \xi ^{\prime }\mid n( \mathbf{q},0)\,\rho(0)
\mid \xi \rangle .
\end{eqnarray}
Inserting this into the expression (\ref{a6}) for the transition
rate we find
\begin{eqnarray}
\dot{W} &= &\lambda ^2\int_0^{\tau_{sc}}  d\tau \,\sum_{\xi ,\xi
^{\prime }}\exp\left[ -i\left( \mathcal{E}_{\xi ^{\prime
}}-\mathcal{E}_\xi \right) t\right] \ \exp \left[ -K\left( \xi
^{\prime }-\xi \right) ^2t\right]
 \nonumber \\
 &&\hspace{2cm}
 \times\langle \xi \mid n(
-\mathbf{q},0)\mid \xi ^{\prime }\rangle \langle \xi ^{\prime
}\mid n(\mathbf{ q},0)\,\rho(0) \mid \xi \rangle \ .
  \label{b-end}
\end{eqnarray}


\section{Decoherence and decrease of transition rate}

Obviously, the decoherence-free limit of this result, i.e. with
$K=0$, corresponds to the conventional result of scattering
theory.

The oscillating factors $\exp\left[ -i\left( \mathcal{E}_{\xi
^{\prime }}-\mathcal{E}_\xi \right) t\right]$  are characteristic
for the 'unitary-type' dynamics  caused by the commutator part
$-i\left[ H,\rho \right]$ of the master equation (\ref{c1}) for
the reduced (or: relevant) density matrix $\rho$. These factors
have the absolute value 1  and do not affect the numerical value
of the transition rate.

On the other hand, the restrictive factors $\exp ( -K\left( \xi
^{\prime }-\xi \right) ^2t) \leq 1$, which are due to the
decoherence,  can be seen to cause a decrease of the transition
rate and thus of the associated cross-section. This can be
illustrated in physical terms as follows:

Let us first assume that the reduced density operator $\rho(0)$
can be chosen to be \textit{diagonal} in the preferred
$\xi-$representation (which corresponds to the usual random phase
approximation at $t=0$). Then each term of Eq.~(\ref{b-end}) is of
the form
\begin{eqnarray}
 \lefteqn{
 \langle\xi| n(-\mathbf{q},0)| \xi ^{\prime }\rangle
\langle \xi ^{\prime }| n(\mathbf{ q},0)\,\rho(0) | \xi \rangle
 =  } \nonumber\\
 & &  \hspace{0.5cm}
  \langle \xi |n( -\mathbf{q},0)\mid \xi ^{\prime }\rangle
\langle \xi ^{\prime }| n(\mathbf{
q},0)|\xi\rangle\langle\xi|\rho(0) | \xi \rangle =  \nonumber \\
 & &  \hspace{2cm}
  |\langle \xi \mid n( -\mathbf{q},0)\mid \xi ^{\prime }\rangle|^2
\langle\xi|\rho(0)| \xi \rangle                \geq 0  \ .
\end{eqnarray}
The last inequality is valid because it holds $\langle\xi|\rho(0)|
\xi \rangle \geq 0$.

 If the assumed diagonal form of $\rho(0)$ would be considered
as being 'too strong', one may note the following: The
exponentials $\exp ( -K\left( \xi ^{\prime }-\xi \right) ^2t)$ due
to decoherence  imply that only terms with $\xi \approx \xi'$
contribute significantly to the transition rate. Thus we may
conclude that, by continuity, all associated terms with $\xi
\approx \xi'$ in Eq.~(\ref{b-end}) should be positive, too. The
further terms with $\xi$ being much different from $\xi'$ can be
positive or negative. But they may be approximately neglected,
since they decay very fast and thus contribute less significantly
to $\dot{W}$.

The main conclusion from the preceding considerations is that the
time average in Eq.~(\ref{b-end}) always decreases the numerical
value of $\dot{W} \equiv W(\tau_{sc})/\tau_{sc}$, due to the
presence of the exponential factors $\exp(-K\left( \xi ^{\prime
}-\xi \right) ^2t) \leq 1$. In other words, the effect of
decoherence during the experimental time window $\tau_{sc}$ plays
a crucial role in the scattering process and leads to an
'anomalous' decrease of the transition rate and the associated
scattering intensity. This result is in line with that  of Ref.
\cite{Schema1}, which investigated the standard expression of the
double differential cross-section of neutron scattering theory by
ad hoc assuming  decoherence of final and initial states of the
scatterer.

Very interesting is also the conclusion that, in the limit of very
slow decoherence ($K\rightarrow 0$), this 'anomaly' disappears,
i.e. the scattering results are expected to agree with
conventional theoretical expectations. This is contrary to the
associated prediction of the theoretical model of
Ref.~\cite{Karlsson1}

\section{Additional remarks}

A related effect (i.e., a shortfall of scattering intensity) was
observed in recent neutron-proton Compton scattering (NCS) and
electron-proton Compton scattering (ECS) experiments in condensed
matter \cite{PRL97,highlights}, in which the experimental
scattering time lies in the sub-femtosecond time scale. This
coincides with the characteristic time of electronic
re-arrangements accompanying the breaking (or formation) of a
chemical bond. Note that in these experiments the energy
transferred to a proton is large enough to break the bond (C--H
and O--H).

Some remarks about the possible selection, definition and/or
physical meaning of the preferred coordinates may be appropriate.
In the case of conventional NCS theory, for example,
 one uses
momentum eigenstates of the scattering particle (e.g. proton)
--- as well as for the neutron --- as the appropriate basis
\cite{Watson}.  In the light of the preceding derivations,
however, one may observe the following: Due to the strong
(Coulomb) interactions  of the scattering proton with its adjacent
particles (electrons, and probably also other nuclei),
$\{|\xi\rangle\}$ can not be one-body states but they should
rather be considered to represent momentum states being strongly
"dressed" (and entangled) with degrees of freedom of adjacent
particles.

Further work will deal with the more general  --- and
experimentally relevant --- case, in which the preferred states
$\{| \xi \rangle\}$ are not eigenstates of the "reduced" energy
 and  Lindbald operators, Eqs. (\ref{b3}). In that case, the
result of Eq.~(\ref{b-end}) will become less simple.

\section{Acknowledgments}
 This work
was partially supported by the EU RT-network QUACS (Quantum
Complex Systems: Entanglement and Decoherence from Nano- to
Macro-Scales).

\end{document}